\documentclass[galley,jgrga]{agu2001}

\usepackage{graphicx}

\authorrunninghead{FRISCH }

\titlerunninghead{HELIOSPHERE-ISM INTERACTION}

\authoraddr{P. C. Frisch,
Department of Astronomy and Astrophysics, University of
Chicago, 5640 S. Ellis Ave., Chicago, IL  60637, USA.
(frisch@oddjob.uchicago.edu) }
\begin{document}
\def\cmtwo{\hbox{cm$^{-2}$}}
\def\kms{\hbox{km s$^{-1}$}}
\def\insitu{\hbox{$in$ $situ$}}
\def\V{\hbox{$V$}}
\def\glong{\hbox{$l$}}
\def\glat{\hbox{$b$}}
\def\cc{\hbox{cm$^{-3}$}}
\def\deeg{\hbox{$^{\rm o}$}}
\def\Bis{\hbox{$B_{\rm IS}$}}
\def\mAA{\hbox{m\AA}}
\def\Lalpha{\hbox{L$\alpha$}}
\def\HI{\hbox{${\rm H^o}$}}
\def\DI{\hbox{${\rm D^o}$}}
\def\NI{\hbox{${\rm N^o}$}}
\def\OI{\hbox{${\rm O^o}$}}
\def\NII{\hbox{${\rm N^+}$}}
\def\CI{\hbox{${\rm C^o}$}}
\def\CIstar{\hbox{${\rm C^{o*}}$}}
\def\CII{\hbox{${\rm C^+}$}}
\def\CIIstar{\hbox{${\rm C^{+*}}$}}
\def\CIV{\hbox{${\rm C^{+3}}$}}
\def\AlII{\hbox{${\rm Al^+}$}}
\def\AlIII{\hbox{${\rm Al^{++}}$}}
\def\MgII{\hbox{${\rm Mg^+}$}}
\def\MgI{\hbox{${\rm Mg^o}$}}
\def\NaI{\hbox{${\rm Na^o}$}}
\def\NCaII{\hbox{$N{\rm (Ca^+)}$}}
\def\CaII{\hbox{${\rm Ca^+}$}}
\def\CaIII{\hbox{${\rm Ca^{++}}$}}
\def\TiII{\hbox{${\rm Ti^+}$}}
\def\SiII{\hbox{${\rm Si^+}$}}
\def\SiIII{\hbox{${\rm Si^{++}}$}}
\def\SiIV{\hbox{${\rm Si^{+3}}$}}
\def\SI{\hbox{${\rm S^+}$}}
\def\SII{\hbox{${\rm S^+}$}}
\def\SIII{\hbox{${\rm S^+}$}}
\def\FeII{\hbox{${\rm Fe^+}$}}
\def\MnII{\hbox{${\rm Mn^+}$}}
\def\ZnII{\hbox{${\rm Zn^+}$}}
\def\NeI{\hbox{${\rm Ne^o}$}}
\def\ArI{\hbox{${\rm Ar^o}$}}
\def\HeI{\hbox{${\rm He^o}$}}
\def\HeII{\hbox{${\rm He^+}$}}
\def\nHeI{\hbox{$n{\rm (He^o)}$}}
\def\nHI{\hbox{$n{\rm (H^o)}$}}
\def\nHII{\hbox{$n{\rm (H^+)}$}}
\def\ntot{\hbox{$n_{\rm H}$}}
\def\NHI{\hbox{$N{\rm (H^o)}$}}
\def\NHeI{\hbox{$N{\rm (He^o)}$}}
\def\NHtot{\hbox{$N{\rm (H^o + H^+)}$}}
\def\NOI{\hbox{$N{\rm (O^o)}$}}
\def\NNI{\hbox{$N{\rm (N^o)}$}}
\def\NHII{\hbox{$N{\rm (H^+)}$}}
\def\NH{\hbox{$N{\rm (H)}$}}
\def\nel{\hbox{$n{\rm (e^-)}$}}
\def\HII{\hbox{${\rm H^+}$}}
\def\nts{\hbox{$n_{\rm TS}$}}
\def\nis{\hbox{$n_{\rm IS,Sun}$}}
\def\Th{\hbox{$T_{\rm H}$}}

%
%

%
\setkeys{Gin}{draft=false}

%
%

%
%

\title{Boundary Conditions of the Heliosphere}
%

%
%


\author{P. C. Frisch}
\affil{Department of Astronomy and Astrophysics, 
University of Chicago, Chicago, Illinois, USA}

%
%


\begin{abstract}

Radiative transfer equilibrium models of nearby interstellar matter
(ISM) are required to determine the boundary conditions of the
heliosphere from astronomical observations of nearby stars.  These
models are also constrained by data on the ISM inside of the solar
system, including pickup ion, anomalous cosmic rays, and \insitu\
\HeI\ data.  The two best ISM models give semi-empirical filtration
factors for H (0.41 - 0.52), He (0.86--1.16), N (0.74--1.11), O
(0.55--0.77), Ne (1.41--2.80), and Ar (0.61--1.76) when observational
uncertainties are included.  Uncertainties in the Ne filtration
factor may result from poorly known interstellar abundances.
These models predict the characteristics of
the ISM outside of the solar system: \nHI=0.20--0.21 \cc, \nel=0.10
\cc, \HII/H=0.29--0.30, and \HeII/He=0.47--0.51.  However, {\it if}
the isotropic 2 kHz emission observed by Voyager (Kurth \&
Gurnett 2003) is formed in the surrounding ISM, an alternate model
(Model 25) is indicated.  The weakly polarized starlight of nearby
stars suggests that the local galactic magnetic field is parallel to
the galactic plane, and the strongest polarization is towards the
upstream direction of the ISM flow, and also (coincidently) near the
ecliptic plane.  Observations of nearby ISM, the radiative transfer
models, and historical $~ ^{10}$Be records provide information on past
variations in the galactic environment of the Sun.
\end{abstract}

%
%

%

\begin{article}

\section{Introduction}

The boundary conditions of the heliosphere are dominated by the
ionization, density, temperature, and magnetic field of the
surrounding interstellar cloud.  The first use of anomalous cosmic ray
(ACR) data to study interstellar ionization compared
ACRs and the \MgI/\MgII\ ratio towards
Sirius, relying on sightline-averaged ionization rates (Frisch 1994).  Full
radiative transfer equilibrium models of interstellar gas within
$\sim$3 pc are now available, predicting ISM properties at both
the heliosphere boundary and averaged over sightlines towards
nearby stars (Slavin \& Frisch 2002, SF02, Frisch \& Slavin 2003, FS03).
Such models provide a tool for evaluating the boundary conditions 
of the heliosphere
when constrained with interstellar matter (ISM) data
inside and outside of the heliosphere.
However, the loss of interstellar neutrals to
charge exchange with interstellar ions in the heliosheath regions 
(``filtration'', Ripken \& Fahr 1983, Izmodenov et al. 1999, Mueller \&
Zank 2002, Cummings \& Stone 2002, CS02) must be considered when 
interpreting the \insitu\ data.  In this paper, the radiative model
predictions for the physical properties of the ISM at the solar
location are used to determine semi-empirical filtration factors,
extending the discussion presented in FS03.  Models consistent with a
possible interstellar origin for the 2 kHz emission observed by
Voyager in the outer heliosphere (Kurth \& Gurnett 2003) are also
discussed, along with evidence for a nearby interstellar magnetic
field and time-variability of the boundary conditions of the
heliosphere.

\section{Why Radiative Transfer Models?}

The ionization of the ISM at the solar location and boundary conditions
of the heliosphere are highly sensitive to
the interstellar cloud properties integrated from the solar location
(SoL) to the cloud surface.  Since the \HI\ column density
to the local cloud surface is log \NHI$<$18 \cmtwo, and  
photons which ionize H ($>$13.6 eV) and He
($>$24.6 eV) are attenuated to 1/e by log \NHI$\sim$17.2 \cmtwo\ and
log \NHeI$\sim$17.7 \cmtwo\ respectively, the \HI\ and \HeI\
densities at the SoL are sensitive to the radiation
attenuation inside the cloud and to cloud geometry.
The measurements which best distinguish the properties of the
surrounding cloud are those which sample partially ionized elements
with first ionization potentials in the range 13--25 eV (e.g. H, He,
O, N, Ne, and Ar), which also constitute the parent population of
pickup ions and anomalous cosmic rays.  

Radiative transfer (RT) models of ISM near the Sun ($<$3 pc) are
required to predict the ionization of the ISM at the SoL.  RT models  have been
constructed using measurements of the interstellar radiation field in
the interval 300--3000 \AA, a modeled conductive interface between
nearby ISM and the hot plasma ($\sim$10$^{6}$ K) interior of the Local
Hot Bubble, and the cloud composition determined from observations
of ISM both inside and outside of the solar system 
(see SF02 and FS03).
Free parameters in these models include the \HI\ column
density towards the cloud surface (\NHI, which is not directly
measured for the sightline used to constrain the models, $\epsilon$ CMa), the soft
X-ray emission from the Local Bubble hot plasma (characterized by
plasma temperature \Th), the interstellar magnetic field strength
($B$, which regulates conduction at the cloud boundary and the extreme
ultraviolet radiation field, Slavin 1989), and cloud density
(\ntot=\nHI+\nHII).  Sightline-averaged values of the
relative ionizations of H and He in the ISM are found from
observations of the \HeI\ ionization edge (504 \AA) towards nearby
white dwarf stars (40--100 pc), which are strong sources of radiation
in this extreme ultraviolet (EUV) wavelength interval (e.g. Kimble et al. 1993, Vallerga
1996).  These data generally sample sightlines with several blended
clouds, and show a wide variation in ionization
(\NHI/\NHeI$\sim$10--15).  The equilibrium models fitting the high He
ionization seen in the EUV data require an excess of radiation in the
500 \AA\ region, when compared to the soft X-ray data, and the
variation of \HI/\HeI\ with cloud depth requires radiative transfer
models (e.g. Cheng \& Bruhweiler 1990, Kimble et al. 1993, Vallerga
1996, SF02).

The initial set of 25 models was pared to seven ``best'' models
(Frisch and Slavin 2003, FS03) by introducing recent astronomical data
indicating gas-phase interstellar abundances of O/H$\sim$400 ppm in
long ($>$700 pc) interstellar sightlines through predominantly diffuse
clouds (Andre et al. 2002), since O and H ionizations are closely
coupled by charge exchange.  
Data on \MgI/\MgII\ and \CIIstar/\CII\
towards $\epsilon$ CMa (Gry \& Jenkins 2001) then indicated that
Models 2, 8, and 18 provide the best match to astronomical data, 
and Model 11 is within uncertainties
(FS03).  FS03 compared these models with He data inside of the solar
system, and concluded Models 2 and 8 are the best models if He is not
filtered in the heliosheath regions.  
(Ionization and density predictions for the full 25 models are available at 
http://astro.uchicago.edu/$\sim$frisch/SlavinFrisch2002\_tables.)
The principal failures of these
two RT models are that predicted \NeI\ densities are $\sim$50\% of \NeI\
densities inferred from PUIs and ACRs, and cloud temperatures are $\sim$30\% 
larger than values inferred from the \HeI\ data.
The filtration predicted by these models is examined below.

\section{Filtration Factors}

Gloeckler \& Geiss (2001, GG01) made an empirical estimate of hydrogen
filtration, $F_{\rm H}= [\nts(H)/\nts(He)]/\phi$, by comparing H and He densities at the termination shock with the interstellar values, $\phi$=\nis(\HI)/\nis(\HeI),
assuming negligible He filtration.
They assumed $\phi$=11.25 based on white dwarf data which show \NHI/\NHeI$\sim$11.25, yielding $F_{\rm H}$= 0.54.  GG01 also derived an O filtration
factor $F_{\rm O}$= 0.62, using \HI\ and \OI\ column densities towards the
stars Procyon and Capella.
(Volume densities are $n$, in units \cc, and column densities for element X are
$N$(X), in units \cmtwo).
The SF02 models show \NHI/\NHeI=8.8--13.6,
consistent with the large variations found for \NHI/\NHeI\ towards
nearby stars, and predict values 
$\phi$=12.1--14.7 at the SoL.
The relative H and He ionizations at the SoL depend on the model
parameters ($\phi$=$\phi$(\ntot,\NHI,$B$,\Th)), since H-ionizing
photons are attenuated more strongly than He-ionizing photons.

Filtration factors can be determined semi-empirically from comparisons
between ionization predictions for the SoL (Table 7 in SF02) and
\insitu\ data on the byproducts of ISM interactions with the solar
system.  Neutral densities at the termination shock are determined
from pickup ion (GG03), anomalous cosmic ray (CS02), and direct He
measurements (Witte et al. 2003).
These data give \nts(\HeI)=0.0145$\pm$0.0015 \cc\ and
\nts(\HI)=0.095$\pm$0.01 \cc.  The H and He filtration factors
determined from the 25 models are shown in Fig. 1a, along with
theoretical predictions of H and He filtration factors.  The seven
models with O/H$\sim$400 ppm are plotted as filled circles, or triangles 
for models 2, 8, 18 which provide the best match to astronomical data.
Among the three models best matching astronomical data, 2 and 8
are most consistent with predictions of negligible He filtration.
The $F_{\rm He}$ predictions for the upstream direction are $F_{\rm He}$=0.98--0.99 (Mueller and Zank, 2002) and
$F_{\rm He}$=0.94 (CS02).  The hydrogen filtration factors are more
uncertain, $F_{\rm H}$=0.4--0.5 (Mueller \& Zank 2002, Ripken \& Fahr
1983, Izmodenov et al. 1999, Gloeckler \& Geiss 2003).

Empirical filtration factors for Ar, O, and N are also shown in
Fig. 1, and in all cases Models 2 and 8 results are consistent with
the predicted range of values (when uncertainties are included).  Model 18 is not consistent with
theoretical calculations suggesting $F_{\rm He} \sim 1$, but is
consistent with Ar, O, and N predicted filtrations.  Models 2 and 8
provide the best fit to the combined interstellar (Mg and C, FS03) and heliospheric
data, and predict similar boundary conditions for the heliosphere
(\nHI$\sim$0.21 \cc, \nel$\sim$0.10 \cc, see Table 1). 
The radiative transfer models are not well constrained using
heliospheric data alone.  The semi-empirical
filtration factors predicted by Models 2, 8 are, respectively:
H (0.46$\pm$0.05, 0.47$\pm$0.05),
He (0.95$\pm$0.09, 1.06$\pm$0.10),
N  (0.91$\pm$0.17, 0.93$\pm$0.18),
O (0.65$\pm$0.10, 0.67$\pm$0.10),
Ne (1.75$\pm$0.34, 2.34$\pm$0.46),
and
Ar (1.10$\pm$0.49, 1.22$\pm$0.554, for Ar/H=3.16e-6, see below).
The filtration factors based on Model 11 are within the range quoted for
Model 2, although the uncertainties differ.

The result that $F_{\rm Ne}$ $>$1 indicates that either the 
interstellar Ne abundance is poorly understood,
or that a heliosheath process enhances \NeI\ (such as charge exchange with \HeI, with
a similar ionization potential).  The
poorly understood extreme ultraviolet radiation field (570--790 \AA, EUV)
ionizes both \NeI\ and \HeI, and since \HeI\ data are consistent with 
model predictions, uncertainties in the EUV radiation field do not appear as
the likely explanation.  The interstellar Ar abundance is uncertain
(Sofia \& Jenkins, 1998), and the abundance Ar/H=3.16e-6 is adopted
here.

As a consistency check, the ratio of the PUI $n$(O)/$n$(N) to the ISM 
$N$(O)/$N$(N) values have
been calculated, and compared to the ratio $F_{\rm O}$/$F_{\rm N}$.
For the ISM, the ratio \NOI/\NNI\ towards $\epsilon$ CMa is used
(SF02, Gry \& Jenkins 2001) (although this is not equal in detail to
$n$(\OI)/$n$(\NI) at the solar location because \NI\ has a higher
photoionization cross section than \OI).  The results are: ($n$(O)/$n$(N)$_{\rm
pui}$)/($N$(O)/$N$(N)$_{\rm ism}$) = 0.70$\pm$0.28, compared to $F_{\rm
O}$/$F_{\rm N}$=0.72$\pm$0.20, as expected.

\section{2 kHz Radio Emission}

Gurnett et al. (1998, 2003) have detected a relatively isotropic
radio-emission signal at 2 kHz in the outer heliosphere with the
plasma wave instruments on board the two Voyager spacecraft; a
possible explanation for the isotropy is emission at the plasma
frequency of the surrounding interstellar gas providing this emission
can enter the heliosphere (e.g. Izmodenov et al. 1999).  
{\it If } this emission has an
interstellar origin, \nel$\sim$0.05 \cc\ and the closest match to this
electron density among the SF02 models would be Model 25
(\nel$\sim$0.0607 \cc) and Model 20 (\nel$\sim$0.0618 \cc).  Model 25
predicts: \nHI=0.21 \cc, \nHeI=0.017 \cc, \HII/H=0.194, \HeII/He=0.341
and T=5,120 K.  Model 20 predicts: \nHI=0.23 \cc, \nHeI=0.018 \cc,
\HII/H=0.183, \HeII/He=0.346 and T=7,200 K.  The relatively low
ionization levels found by these models results from the absence of a
conductive interface on the cloud, which is characterized by the
magnetic field parameter $B$=0.  Model 25 yields predicted filtration
factors of: 
$F_{\rm H}$=0.45$\pm$0.05, 
$F_{\rm He}$=0.85$\pm$0.08,
$F_{\rm N}$=0.55$\pm$0.11, 
$F_{\rm O}$=0.39$\pm$0.06, 
$F_{\rm Ne}$=1.16$\pm$0.23, and 
$F_{\rm Ar}$=0.83$\pm$0.37.  
GG03 have found that Model 25 offers a self-consistent solution when
heliosphere models of filtration are compared to the PUI data.
However, if Model 25 is correct further study is required to
understand the local ISM towards $\epsilon$ CMa, since the Model 25
predictions for interstellar abundances do not match the local ISM
observations towards $\epsilon$ CMa.

\section{Interstellar Magnetic Field }

The interstellar magnetic field (\Bis) immediately outside of the
heliosphere has not been directly measured.  However, observations of
weakly polarized light from nearby ($<$50 pc) stars show a patch of
interstellar dust centered on galactic coordinates
$l$=350\deeg$\rightarrow l$=20\deeg, $b$=--40\deeg$\rightarrow
b$=--5\deeg, with extinction $A_{\rm V} \sim$0.01 mag (or
$N$(H)$\sim$2 x 10$^{19}$ \cmtwo) and galactic field directed towards
$l \sim$70\deeg\ (Tinbergen 1982, Frisch 1990).  Fig. \ref{fig:pol}
shows the location of the stars (in galactic coordinates), and the
polarization vectors, in the region where the weak polarization is
most evident, with the plane of the ecliptic superimposed.  Also
plotted are the radio emission sources detected by the Voyager
spacecraft which are consistent with an origin near the heliosphere
nose and follow roughly the galactic plane (Kurth \& Gurnett 2003).
The weak polarization is most evident near the plane of the ecliptic
and coincides with the upstream direction of the cluster of
interstellar clouds flowing past the Sun (which is shown in Fig. 2
plotted in the Local Standard of Rest, after correction for solar
motion).  These polarizing dust grains are predominantly within 5 pc
of the Sun, or closer (Frisch 1990).  Although interstellar grain
alignment mechanisms are poorly understood, popular grain alignment
models yield polarization vectors parallel to the interstellar
magnetic field (Lazarian \& Cho 2002).  The same classical
interstellar dust grains which polarize optical radiation also pile up
in the heliosheath regions as grains are deflected around the
heliosphere (grain radius $\sim$0.1 $\mu$m, Frisch et al. 1999).  Rand \& Lyne (1994) compared pulsar
dispersion and rotation measures towards pulsars spaced over several
kpc, to derive the strength of the ordered component of the magnetic
field near the Sun, $B$=1.4$\pm$0.2 $\mu$G, directed towards galactic
longitude $l =88\deeg \pm$5\deeg; this field is consistent with the
field traced by Tinbergen's data.  Once the production mechanisms for
the grain alignment and for the radio emission sources are understood,
it may produce a deeper understanding of the asymmetries in the
heliosphere nose region due to the $\sim$60\deeg\ tilt between the
ecliptic plane and interstellar magnetic field.

\section{Variability in the Boundary Conditions of the Heliosphere.}

Simple geometrical assumptions indicate the Sun
left the Local Bubble interior and entered the surrounding interstellar 
cloud within the past several thousand years (Frisch 1994).
More precise estimates of the variability of the boundary
conditions of the heliosphere result from combining densities from
the best RT models with observations of nearby stars.

Models 2 and 8 give a local ISM density of
\nHI$\sim$0.21 \cc, which when combined with the column density to the
cloud surface (\NHI=6.5 10$^{17}$ \cmtwo) give a distance to the cloud
surface of $\sim$0.97 pc.  This distance is traversed in $\sim$37,000
years by a cloud at the LIC velocity.  Since the local ISM towards
$\epsilon$ CMa and $\alpha$ CMa (Sirius) consists of two cloudlets
(Gry \& Jenkins 2001, Lallement et al. 1994), and the second
blue-shifted cloud has a higher velocity relative to the Sun than the
LIC, the time at which the Sun encountered the LIC must be shorter than
37,000 years.  Frisch (1997) attributed spikes in the $~ ^{10}$Be ice
core record 33 kyr and 60 kyr ago to encounters with $\sim$0.1 pc wide
magnetic structures, comoving in the local ISM, which confine low
energy galactic cosmic rays.  These structures would now be $\sim$0.87
pc, and $\sim$1.6 pc distant from the Sun in the downstream direction
(e.g. towards $\epsilon$ CMa, $\alpha$ CMa).  The most recent $~
^{10}$Be spike may be consistent with the solar entry into the local
ISM; alternatively it may correlate with the Sun's entry into the LIC while the earlier event may signal
entry into the blue-shifted cloud seen towards CMa.  Prior to
that, the Sun was immersed in the hot ($10^4$ K) plasma interior to
the Local Bubble where an extended heliosphere (heliopause radius 300
au in the nose direction) is predicted (Frisch et al. 2003).  
(Civilization developed as the Sun emerged from the Local Bubble interior,
where it resided for several millions of years.)

Looking to the future, at least three interstellar clouds are found
within 5 pc of the Sun in the upstream direction, the Local
Interstellar Cloud (LIC) where the Sun is now located, the
``G-cloud'', and the Apex Cloud (Frisch 2003).  The velocity vector of
each cloud is consistent with the cloud being located in front of the
nearest star $\alpha$ Cen, although the G-cloud is the best fit to
observations (e.g. Lallement et al. 1995, Linsky \& Wood 1996) and the LIC 
would have to have a small velocity gradient
($\sim$0.7 km/s) towards the star.  If the G-cloud is the next cloud
to be encountered by the Sun, the higher heliocentric velocity (--29.1
km/s, factor of $\sim$1.1 ) and neutral density ($>$5 \cc, factor of $>$24)
than for the LIC indicates a dramatic heliosphere shrinkage if the
other properties (e.g. ionization) are similar to the LIC (Frisch
2003, Zank \& Frisch 1999).  If, in contrast, the Apex Cloud is the
next cloud encountered, the larger heliocentric velocity (--35.1 \kms)
also indicates a smaller heliosphere, although the other properties of
this cloud are unknown.  Variations of the boundary conditions of the
heliosphere are thus expected on timescales of less than
37,000--45,000 years if either the G-cloud or Apex Cloud is foreground
to $\alpha$ Cen.

\section{Discussion}

The primary results of this paper are estimates of filtration
factors for H, He, N, O, Ne, and Ar, based on radiative transfer models of the local ISM from SF02, and FS03.
Two radiative transfer equilibrium models of 
ISM within $\sim$3 pc of the Sun (models 2 and 8) yield predictions which
match observations of ISM
inside (He, PUIs, ACRs) and outside (gas within $\sim$3 pc towards $\epsilon$ CMa)
of the solar system,
for an assumed interstellar gas-phase O abundance of O/H=400 ppm (FS03).
These models yield
semi-empirical predictions for He, H, O, N, Ar filtration factors
which are consistent with theoretical predictions, within uncertainties, and
predict interstellar densities \nHI$\sim$0.20 \cc, \nel=0.10 \cc\ at the
solar location.  
However, there are unexplained differences between the predictions 
of Models 2 and 8 and observed pickup ion Ne densities and the cloud temperature.

If the astronomical constraints on the radiative transfer models
are relaxed, alternate models become plausible.  For example,
if the 2 kHz radio emission observed by Voyager 1, 2
is from the surrounding ISM, 
the low electron density of Model 25 (\nel$\sim$0.06 \cc) is a better match
although predicted interstellar abundances are unrealistic. 
For Model 25 to be correct the radiation field intensity versus cloud
geometry requires further study.
The simple interstellar cloud geometry assumed for the underlying radiative 
transfer models may also prove inadequate.
If the astronomical constraints prove incorrect (e.g. the assumed
interstellar gas-phase oxygen abundance), then the
selection of the best models to describe the boundary conditions of the
heliosphere may also change.  

As better data on the ISM inside and outside of the
heliosphere become available, yielding more accurate boundary 
conditions for the heliosphere, heliosphere modeling will predict
more reliable filtration factors for comparison with semi-empirical
filtration factors utilizing interstellar modeling.

The strength of the interstellar magnetic field outside of the
heliosphere remains uncertain, but both low frequency radio
emission events and the polarization of the light of nearby stars
suggest the field orientation is parallel to the galactic plane
and not the ecliptic plane, introducing a north/south heliosphere asymmetry.

The second important conclusion of this paper is that
the boundary conditions of the heliosphere change with time,
and on timescales of $<$10$^4$ years,
indicating past and future variations in heliosphere dimensions and properties.

%
%

\begin{acknowledgments}
The author acknowledges research support from NASA grants NAG5-6405,
and NAG5-11005, and NAG5-8163.
The author would like to thank the
referees for helpful comments which improved this paper.
\end{acknowledgments}

%
%

\newpage
\section{References}

\begin{itemize}

\item
Andre, M., C.~Oliveira, J.~C. Howk, R.~Ferlet, J.~M. Desert,
  G.~Hebrard, S.~Lacour, A.~Lecavelier des Etangs, A.~Vidal-Madjar,
  and H.~W. Moos, Oxygen gas phase abundance revisited, {\it \apj\/}, p.
  submitted, 2002.

\item
Cheng, K. and Bruhweiler, F.~ C.,
Ionization Processes in the Local Interstellar Medium - Effects of the
Hot Coronal Substrate, {\it \apj} {\it 364}, 573--581, 1990.

\item
Cummings, A.~C., E.~C. Stone, and C.~D. Steenberg, Composition of
  Anomalous Cosmic Rays and Other Heliospheric Ions, {\it \apj\
  578\/}, 194--210, 2002.

\item
Frisch, P.~C., Characteristics of the local interstellar medium, in {\it
  Physics of the Outer Heliosphere\/}, pp. 19--22, 1990.

\item Frisch, P. C., Journey of the Sun, http://xxx.lanl.gov/astro-ph/970523, 1997.


\item
Frisch, P.~C., Morphology and ionization of the interstellar cloud
  surrounding the solar system, {\it Science,  265\/}, 1423, 1994.

\item
Frisch, P.~C., Local Interstellar Matter: The Apex Cloud, {\it \apj\
  in press}, 2003.

\item
Frisch, P.~C., and J.~D. {Slavin}, Chemical Composition and Gas-to-Dust Mass
  Ratio of the Nearest Interstellar Matter, {\it \apj\   in press \rm\/},
  2003.

\item
Frisch, P.~C., J.~M. Dorschner, J.~Geiss, J.~M. Greenberg, E.~Gr\"un,
  M.~Landgraf, P.~Hoppe, A.~P. Jones, W.~Kr\"atschmer, T.~J.
  Linde, G.~E. Morfill, W.~Reach, J.~D. Slavin, J.~Svestka, A.~N.
  Witt, and G.~P. Zank, Dust in the Local Interstellar Wind, {\it
  \apj\   525\/ }, 492--516, 1999.

\item
Frisch, P.~C., L.~Grodnicki, and D.~E. Welty, The Velocity Distribution
  of the Nearest Interstellar Gas, {\it \apj\/} {\it 574\/}, 834--846, 2002.

\item 
Frisch, P.~C., H. R.~ Mueller, G. P.~ Zank, and C.~ Lopate, Galactic
Environment of the Sun and Stars: Interstellar and Interplanetary
Material, in {\it Astrophysics of Life}, Eds. M. Livio, I. N. Reid,
and W. B. Sparks (Cambridge: Cambridge University Press), 2003. 

\item
Gloeckler, G., and J.~Geiss, Composition of the Local Interstellar Cloud
  from Observations of Interstellar Pickup Ions, in {\it AIP Conf. Proc. 598:
  Joint SOHO/ACE workshop "Solar and Galactic Composition"\/}, pp. 281--+,
  2001.

\item
Gloeckler, G., and J.~Geiss, Composition of the Local Interstellar Cloud
  as Diagnosed with Pickup Ions , {\it Adv. Space Res, in press\/}, 2003.

\item
Gry, C., and E.~B. Jenkins, Local clouds: Ionization, temperatures,
  electron densities and interfaces, from GHRS and IMAPS spectra of epsilon
  Canis Majoris, {\it \aap\/} 367, 617--628, 2001.

\item
Izmodenov, V.~V., J.~Geiss, R.~Lallement, G.~Gloeckler, V.~B.
  Baranov, and Y.~G. Malama, Filtration of interstellar hydrogen in the
  two-shock heliospheric interface: Inferences on the local interstellar cloud
  electron density, {\it \jgr\/} {\it 104\/}, 4731--4742, 1999.

\item
Kimble, R. A., Davidsen, A. F., Blair, W. P., 
        Bowers, C. W., Dixon, W. V. D., Durrance, S. T.,
        Feldman, P. D., Ferguson, H. C., Henry, R. C.,
        Kriss, G. A., Kruk, J. W., Long, K. S., Moos, H. W.,
        Vancura, O.,
Extreme Ultraviolet Observations of G191-B2B and the 
Local Interstellar Medium with the Hopkins Ultraviolet Telescope,  
{\it \apj} {\it 404}, 663--672, 1993.

\item
Kurth, W.~S., and D.~A. Gurnett, On the source location of low-frequency
  heliospheric radio emissions, {\it J. Geophys. Res.\/}, {\it this issue\/},
  000, 2003.

\item
Lallement, R., Bertin, P., Ferlet, R., Vidal-Madjar, A., and Bertaux, J.,
GHRS observations of Sirius-A I. Interstellar clouds toward Sirius 
and Local Cloud ionization, {\it \aap} {\it 286}, 898--908, 1994.

\item
Lallement, R., R.~Ferlet, A.~M. Lagrange, M.~Lemoine, and A.~{Vidal-Madjar},
  Local cloud structure from HST-GHRS, {\it \aap\/} {\it 304\/}, 461--474,
  1995.

\item
Linsky, J.~L., and B.~E. Wood, The alpha Centauri line of sight: D/H
  ratio, physical properties of local interstellar gas, and measurement of
  heated hydrogen (the `hydrogen wall') near the heliopause, {\it \apj\/} {\it
  463\/}, 254--270, 1996.

\item
Lazarian, A., and J.~Cho, Polarized Foreground Emission from Dust: Grain
  Alignment and MHD Turbulence, {\it Elseview Science\/}, submitted, 2002.

\item
Mueller, H., and G.~P. Zank, Modeling Heavy Ions and Atoms throughout
  the Heliosphere, in {\it "Solar Wind 10"\/}, 2002.

\item
Rand, R.~J., and A.~G. Lyne, New Rotation Measures of Distant Pulsars in
  the Inner Galaxy and Magnetic Field Reversals, {\it \mnras\/} {\it 268\/},
  497--+, 1994.

\item
Ripken, H.~W., and H.~J. Fahr, Modification of the local interstellar gas
  properties in the heliospheric interface, {\it \aap\/} 1983.

\item
Slavin, J.~D., Consequences of a Conductive Boundary on the Local Cloud, 
{\it \apj\/} {\it 346\/}, 718--727 1989.

\item
Slavin, J.~D., and P.~C. Frisch, The Ionization of Nearby Interstellar
  Gas, {\it \apj\/  565} 364--379, 2002.

\item
Sofia, U.~J., and E.~B. Jenkins, Interstellar medium absorption profile
  spectrograph observations of interstellar neutral argon and the implications
  for partially ionized gas, {\it \apj\/ 499} 591, 1998.

\item
Tinbergen, J., Interstellar polarization in the immediate solar neighborhood,
  {\it \aap\/} {\it 105\/}, 53--64, 1982.

\item
Vallerga, J., Observations of the Local Interstellar Medium with
the EUVE,  {\it Space Science Reviews \/}, {\it 78}, 277--288, 1996.

\item
Witte, M., M.~Banaszkiewicz, H.~Rosenbauer, and D.~McMullin, Kinetic parameters
  of interstellar neutral helium: Updated results from the {ULYSSES/GAS}
  instrument, {\it Adv. Space Res, in press\/}, 2003.

\item
Zank,  G.~P. and Frisch, P.~C., Consequences of a Change in the Galactic 
Environment of the Sun,  { \it \apj\  \it 518}, 965--973, 1999.

\end{itemize}
\end{article}
\newpage

%
%
%

\begin{figure}
\begin{center}
\includegraphics[angle=0, width=35pc]{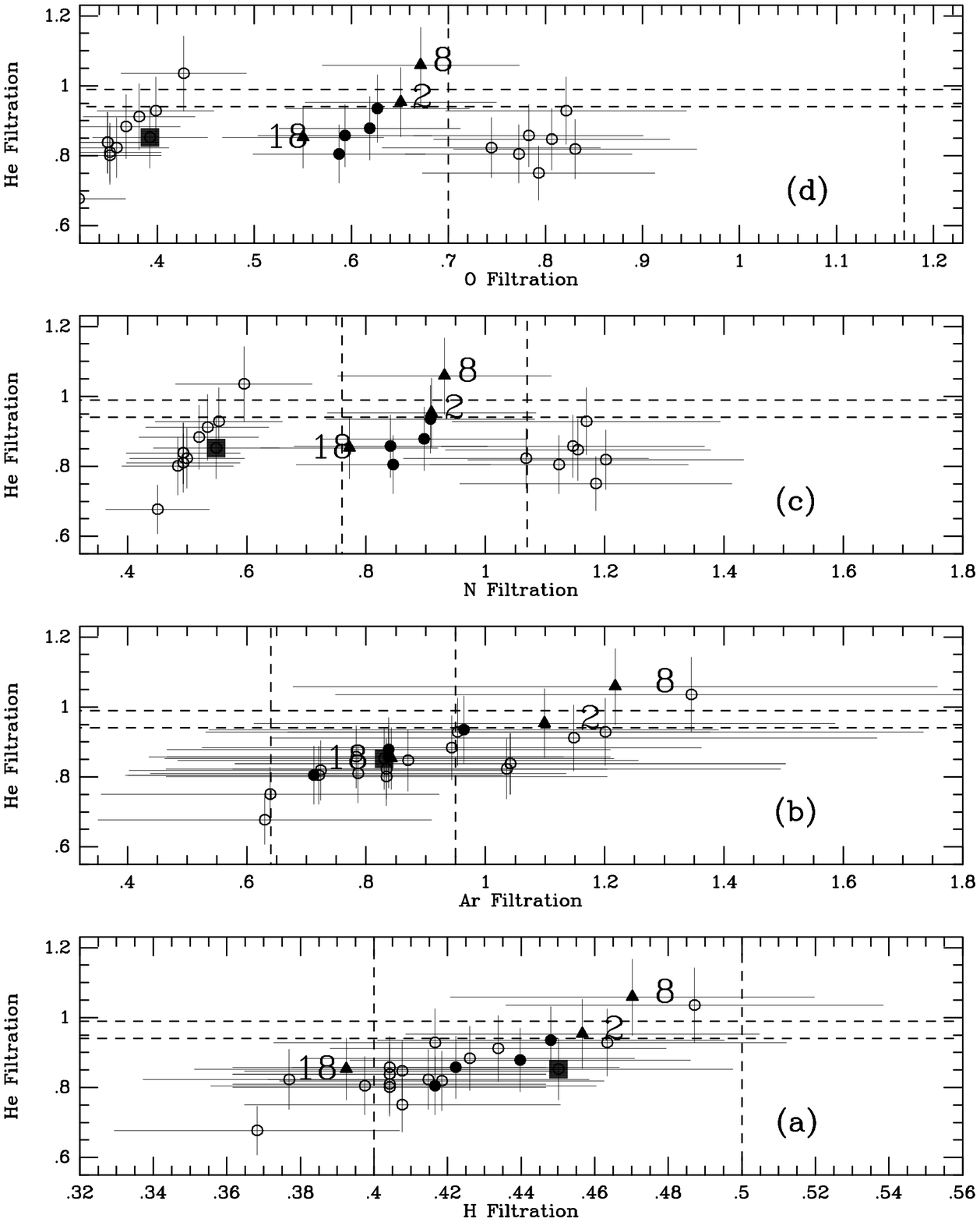}
\end{center}
\caption{
Empirical filtration factors for 
H (a), Ar (b), N (c), and O (d) are plotted against $F_{\rm He}$, as
derived from RT models (SF02, Table 7) compared to
PUI, ACR, and He data (including data uncertainties).
Filled circles and triangles are ISM models with O/H$\sim$400, 
and the labeled triangles are nos. 2, 8, 18
which match local ISM Mg$^{\rm +}$/Mg$^{\rm o}$ and
C$^{\rm +}$/C$^{\rm +*}$ data.  The box shows Model 25 
(consistent with a low electron density).
Dashed lines are theoretical filtration factors.
\label{fig:HHe}
}
\end{figure}

\begin{figure}
\noindent\includegraphics[angle=270,  width=20pc]{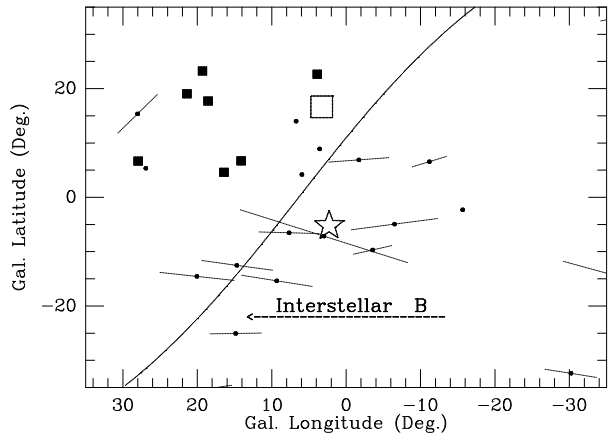}
\caption{Plot of an indicator of the nearby interstellar magnetic
field, in galactic coordinates.  The bars show the direction of the
electric vector polarization, which is parallel to the interstellar
magnetic field direction, for several nearby stars (Tinbergen 1982).
The arrow shows the likely direction of the interstellar magnetic
field near the Sun based on the polarization and pulsar (Rand \& Lyne
1994) data.  The curved line is the ecliptic plane.  The observed
region of maximum polarization is near the ecliptic plane, but is also
in the upstream direction (as referred to the ``LSR'' velocity rest frame of 
stars close
to the Sun, Frisch et al. 2002) of the local flow of ISM past the Sun (star).  
The heliosphere nose
direction, in the heliocentric rest frame, is located at the box.  For
comparison, the sources determined for the $\sim$3 kHz radio emissions
detected by Voyager are plotted as filled boxes (Kurth \& Gurnett
2003).  \label{fig:pol}}
\end{figure}

\begin{table}
\caption{ISM Physical Properties at the Heliosphere\tablenotemark{a}}
\begin{flushleft}
\begin{tabular}{lcc}
\tableline
\multicolumn{1}{c}{Quantity}& \multicolumn{2}{c}{Models} \\
 & 2 & 8 \\
\tableline
\multicolumn{3}{c}{{\it Assumed Model Parameters }} \\
$n_{\rm H}$ (\cc) \tablenotemark{b}& 0.273 & 0.273 \\
log $T_{\rm h}$ (K) & 6.0  & 6.1 \\
$B_{\rm o}$ ($\mu$G) & 5.0 & 5.0 \\
$N_{\rm H^o} $ (10$^{17}$ \cmtwo)\tablenotemark{c} & 6.5 & 6.5 \\
\multicolumn{3}{c}{{\it Predicted Quantities at Solar Location }} \\
$n$(\HI) (\cc) & 0.208 & 0.202 \\
$n$(\HeI) (\cc) & 0.015 & 0.014 \\
$n$(e) (\cc) & 0.098 & 0.101 \\
$\chi$(H)\tablenotemark{d} & 0.287 & 0.300 \\
$\chi$(He)\tablenotemark{d} & 0.471 & 0.511 \\
T (K) & 8,230 & 8,480 \\
$n$(\NI)/\nHeI & 5.64e-4 & 6.12e-4 \\
$n$(\OI)/\nHeI & 5.35e-3 & 5.76e-3 \\
$n$(\NeI)/\nHeI & 3.86e-4 & 2.37e-4 \\
$n$(\ArI)/\nHeI\tablenotemark{e} & 1.21e-5 & 1.21e-5 \\
\multicolumn{3}{c}{{\it Predicted Quantities towards $\epsilon$ CMa}} \\
log $N_{\rm H }$ (\cmtwo)\tablenotemark{b,c} & 18.03 & 18.02 \\
$N$(\HI)/$N$(\HeI)\tablenotemark{c}  & 11.63 & 12.74 \\
\tableline
\end{tabular}
\end{flushleft}
\tablenotetext{a} {From (Slavin \& Frisch 2002) and (Frisch \& Slavin 2002).} 
\tablenotetext{b} {$n_{\rm H } = n$(\HI)+$n$(\HII) \cc;
$N_{\rm H } = N$(\HI)+$N$(\HII) \cmtwo.} 
\tablenotetext{c}  {$N$(H$^o$) is the \HI\ column density 
between the Sun and cloud surface, 
etc., where the cloud is the sum of the two nearby cloudlets ($d$$<$3 pc) 
observed towards $\epsilon$ CMa (Gry \& Jenkins 2001).} 
\tablenotetext{d} {$\chi$(H), $\chi$(He) are the ionized fractions of H, He, respectively.}  
\tablenotetext{e} {The B-star Ar abundance of 3.16 ppm is used here (from Sofia \& Jenkins 1998).}
\end{table}

\end{document}